# Mesoscopic Fluctuations of Elastic Cotunneling in Coulomb Blockaded Quantum Dots

S. M. Cronenwett, S. R. Patel, C. M. Marcus
*Department of Physics, Stanford University, Stanford, California 94305-4060*

K. Campman and A. C. Gossard
*Materials Department, University of California at Santa Barbara, Santa Barbara, California 93106*

We report measurements of mesoscopic fluctuations of elastic cotunneling in Coulomb blockaded quantum dots. Unlike resonant tunneling on Coulomb peaks, cotunneling in the valleys is sensitive to charging effects. We observe a larger magnetic field scale for the cotunneling (valley) fluctuations compared to the peaks, as well as an absence of “weak localization” (reduced conductance at B=0) in valleys. Cotunneling fluctuations remain correlated over several valleys while peak conductance correlation decreases quickly.

The Coulomb blockade (CB) in small metallic and semiconductor structures provides a system in which transport is dominated by the effects of electron-electron interactions [1, 2]. At low temperature, transport in small systems is also influenced by quantum interference, leading to mesoscopic fluctuations with universal statistical features [3, 4]. Recent theoretical work has provided a detailed understanding of mesoscopic fluctuations for the case of non-interacting electrons and has called attention to the connection between the universal statistics and quantum chaos [5-8]. Coulomb blockaded quantum dots provide a useful system to move beyond the non-interacting picture, yet remaining relatively simple: interactions can often be treated in terms of a single charging energy.

The dominant experimental signature of the CB is the appearance of narrow peaks in conductance as an external gate voltage is swept, changing the potential of the dot. Conduction is suppressed between peaks when the temperature, $T$, and voltage bias, $V_{SD}$, are less than the energy $E_C = e^2/C$ required to add one electron to the dot. In the quantum regime, $(kT, V_{SD}) < \Delta << E_c$, (where $\Delta = 2\pi\hbar^2 / m^* A$ is the mean level spacing and $A$ is the dot area) transport on CB peaks is mediated by resonant tunneling and is insensitive to charging effects. This has been demonstrated, for instance, in recent measurements of peak height statistics [9, 10] which gave excellent agreement with single-particle random matrix theory (RMT) [11-13]. In contrast, conduction *between* CB peaks is mediated by so-called cotunneling, which is sensitive to charging [14-17]. At low temperatures, the dominant cotunneling process is elastic and, like the peaks, exhibits interference effects in the form of random but repeatable conductance fluctuations [18].

In this Letter, we report the first measurements of mesoscopic fluctuations of elastic cotunneling in the valleys between CB peaks. The quantum dots used are ballistic, with a "chaotic" shape allowing comparison to a universal theory [18] and make use of shape-distorting electrostatic gates (Fig. 2(b), inset) to allow ensemble statistics to be gathered on a single device. We observe an increase in the characteristic magnetic field scale of conductance fluctuations in the valleys compared to peaks in the CB regime and find that valleys do not exhibit the decrease in average conductance at B=0 ("weak localization") that is seen in peaks. We also compare cross-correlations of neighboring peaks and valleys, finding longer correlation among valleys. This effect is explained in terms of the number of quantum levels participating in transport.

In the valleys between CB peaks, conductance is mediated by two mechanisms: elastic and inelastic cotunneling [16] . Each contributes to the average conductance,

$$\left\langle G_{el}\right\rangle = \frac{\hbar G_l G_r \Delta}{4\pi e^2}\left(\frac{1}{E_e}+\frac{1}{E_h}\right),\ \left\langle G_{in}\right\rangle = \frac{\hbar G_l G_r \pi}{3e^2}(kT)^2\left(\frac{1}{E_e}+\frac{1}{E_h}\right)^2, \qquad (1)$$

where $E_{e(h)}$ is the difference between the Fermi energy in the leads and the next available state for electron (hole) tunneling (i.e. $E_e + E_h = E_c$, and $E_e = E_h = E_c/2$ in the middle of the valley, see Fig. 2(c), inset). At low temperatures, $kT < \sqrt{E_c\Delta}$, conductance is dominated by the elastic mechanism, consisting of virtual tunneling of charge over an energy barrier of height ~$E$, the smaller of ($E_e$, $E_h$). Near the valley center, where $E \sim E_c/2 >> \Delta$, virtual tunneling takes place through a large number $\sim E/\Delta$ of levels. Conduction through each level fluctuates randomly with external parameters, resulting in mesoscopic cotunneling fluctuations that do not average away even for large $E/\Delta$ [18].

Experimentally, one can distinguish resonant tunneling fluctuations (on peaks) from cotunneling fluctuations (in valleys) by the characteristic magnetic field scale, $B_c$, defined as the width of the autocorrelation $C_{i,i}(\Delta B)$, where $C_{i,j}(\Delta B) = \left\langle \tilde{g}_i(B)\tilde{g}_j(B+\Delta B)\right\rangle_B / \left(\sqrt{\mathrm{var}\,\tilde{g}_i}\sqrt{\mathrm{var}\,\tilde{g}_j}\right)$, and $\tilde{g} = G - \langle G\rangle_B$. In particular, $B_c^{valley}$ is the field required to pass roughly one flux quantum, $\phi_0 = h/e$, through a typical area difference accumulated by chaotic trajectories in the time $\tau \sim h/E$ limited by virtual tunneling at energy $E$, giving

$$B_c^{valley} = \kappa(\phi_0 / A)\sqrt{E/E_T}, \qquad (2)$$

where $E_T = \hbar v_F / A^{1/2}$ is the ballistic Thouless energy and $\kappa$ is a device-dependent geometrical factor [18]. Equation (2) applies when most virtual trajectories are fully chaotic, $E < E_T$. In the opposite case, $E > E_T$, $E_T$ replaces $E$, giving $B_c^{valley} \sim \kappa\phi_0/A$. For comparison, the characteristic field scales of conductance fluctuations on CB peaks and in open quantum dots are

$$B_c^{peak} = \kappa(\phi_0 / A)\sqrt{\Delta/E_T},\ B_c^{open} = \kappa(\phi_0 / A)\sqrt{\Gamma_{tot}/E_T}, \qquad (3)$$

where $\Gamma_{tot} = \Gamma_l + \Gamma_r + \Gamma_\varphi$ is the total broadening due both to escape ($\Gamma_{l,r} = (h\Delta/\pi e^2)G_{l,r}$) and dephasing [7, 19, 20]. One expects ($B_c^{peak}, B_c^{open}$) < $B_c^{valley}$ from Eqs. (2) and (3) since $(\Delta, \Gamma_{tot}) < E \sim E_c / 2$. Equation (2) further implies that $B_c{}^{valley}$ will have a maximum at mid-valley, where $E = E_c / 2$, and decrease to match $B_c{}^{peak}$ as the peak is approached and $E \to 0$. Making the approximation $\langle G_{el} \rangle \propto (E_e{}^{-1} + E_h{}^{-1}) \propto \sim E^{-1}$ and noting that $B_c{}^{valley} \propto \sqrt{E}$ further implies that $B_c{}^2 \langle G_{el} \rangle$ ~ constant.

We report measurements for three quantum dots defined using Cr/Au electrostatic gates on GaAs/AlGaAs heterostructures. Important parameters for the dots are given in Fig. 4(d). The dots use similar designs with adjustable point contacts and two or three shape-distorting gates as shown in Fig. 2(c). Measurements were made in a dilution refrigerator using an ac voltage bias with $V_{SD}$ ~ 5 μV (< Δ, *kT*) at 13Hz. The electron temperature, *T* ~ 100 mK, measured from CB peak widths, easily satisfies the requirement $kT < \sqrt{E_c \Delta}$ ~ 600 mK. Conductance fluctuation statistics were measured by rastering over gate voltage, $V_g$, and field, *B*, to yield a 2D scan of conductance, as seen in Fig. 1(a), then finding the maxima of peaks from $cosh^{-2}$ fits [21] and the minima of valleys from parabolic fits. Random conductance fluctuations are seen in the peaks, valleys, and all interpolated paths (see Fig. 1(b)). The field scale of fluctuations is defined by the condition $C_{i,i}(0.325 B_c) = 0.82$. This somewhat awkward definition was chosen to coincide with Ref. [18] while allowing $B_c$ to be measured in the universal region, $C_{i,i}(\Delta B \sim 0) \sim 1$, where presumably nonuniversal features of the dot geometry such as short trajectories are not important. The periodic change in $B_c$ from a maximum in the valley to a minimum on peak can be seen in the width of the $C_{i,i}(\Delta B)$ color scale, Fig. 1(c), or directly from the autocorrelations shown in Fig. 1(d). We note that the expected difference in the *functional forms* of $C_{i,i}(\Delta B)$ for peaks versus valleys [12, 13, 18] is not resolvable in the data.

The ensemble-averaged field scale, $\overline{B_c}$, and conductance, $\overline{\langle G \rangle_B}$, were obtained from statistically independent peak–valley–peak data sets, sampled by changing the voltage applied to one of the shape-distorting gates. Figure 2(a) shows $\overline{B_c}$ and $\overline{\langle G \rangle_B}$ for 31 peaks and 22 valleys from dot 1, estimated to contain ~14 statistically independent data sets. The measured values, $\overline{B_c}^{valley}$ = 6.3 ± 0.3 mT and $\overline{B_c}^{peak}$ = 4.0 ± 0.2 mT, give a ratio $\overline{B_c}^{valley} / \overline{B_c}^{peak}$ ~ 1.6, consistent with ratios of 1.4 - 1.8 for the other dots. While the general behavior of $B_c$ in valleys compared to peaks agrees with theoretical expectations, the ratio is smaller than expected from Eqs. (2) and (3). In particular, one would expect $\overline{B_c}^{valley} / \overline{B_c}^{peak}$ ~ (18mT /4.6 mT) ~ 4 for dot 1. This discrepancy is found in all of the dots and is not understood at present. Figure 2(b) shows the approximately linear dependence of $\overline{B_c}^{-2}$ on $\overline{\langle G \rangle_B}$ for the data in Fig. 2(a). We note, however, that this relation assumes only a single carrier—either electrons or holes—and so can only be applied near the valley bottom or peak top.

To show that $B_c^{valley}$ is enhanced due to charging effects, it is useful to compare $B_c$ for both peaks and valleys to the characteristic field $B_c{}^{open}$ for open dots ($G_{dot} > e^2/h$). Ensemble-averaged field scales, $\overline{B_c}^{open}$ (using the definitions of $C_{i,i}(\Delta B)$ and $B_c$ above), obtained from ~30 statistically independent conductance traces at three different lead conductances in dot 1 are shown in Fig. 2(c) along with $\overline{B_c}^{peak}$ ($\overline{\langle G \rangle_B}$ ~ 0.4 $e^2/h$) and $\overline{B_c}^{valley}$ ($\overline{\langle G \rangle_B}$ ~ 0.05 $e^2/h$) for the tunneling regime. These data show that $\overline{B_c}^{open}$ converges to $\overline{B_c}^{peak}$ as the dot becomes isolated, while $\overline{B_c}^{valley}$ is considerably larger than either. This supports the observation that the characteristic (single-particle) energy scales for transport though open dots and for resonant tunneling, Γ and Δ, coincide at the onset of blockade, and that both are smaller than *E* (set by classical charging) which determines $\overline{B_c}^{valley}$.

It is known theoretically [11, 22] and from recent experiments [9, 10] that the normalized on-peak conductance, $\bar{g}(B) = \overline{G(B)/\langle G \rangle_{B\neq 0}}$, is lower when time-reversal symmetry is obeyed (i.e. at B=0), in analogy to weak localization in open dots [23, 24] and 1D and 2D disordered conductors [25]. In contrast, elastic cotunneling in the valleys is not expected to show weak localization—that is, $\bar{g}_{valley}$ should not be suppressed at B=0 [18]. We have investigated the change in average conductance, $\delta g = (\bar{g}(B > B_c) - \bar{g}(B = 0))$, for 81 independent pairs of peaks and valleys measured in dot 3. As shown in Fig. 3, we find $\delta g_{peak} = 0.14$ for the peaks, somewhat smaller than the RMT result, $\delta g_{peak} = 1/4$ [11, 22]. Since the RMT calculation assumes $\Gamma << kT << \Delta$ while the measurement has $\Gamma \sim 0.7\Delta$ and $kT \sim \Delta$, it is reasonable that theory should overestimate the measured value. For the valleys, we find that $\bar{g}_{valley}$ lacks any significant dip on a field scale of $B_c$ (~ 8 mT for dot 3) around B=0 (Fig. 3(b)), in agreement with theory [18]. We note that averages such as $\bar{g}(B)$ are difficult to measure in the CB regime because, unlike in open dots, fluctuations in $g$ are on the order of $g$ itself [11, 18, 22], as illustrated in Fig. 3.

Finally, we investigate correlations between neighboring peaks and valleys as a function of separation (in units of peak spacing), $\Delta N$. Ensemble-averaged cross-correlations, $\overline{C(\Delta B, \Delta N)} \equiv \overline{C_{i,i+\Delta N}(\Delta B)}$, for 5 peaks and valleys are shown in Fig. 4. The maximum of $\overline{C(\Delta B, \Delta N)}$ at $\Delta B = 0$ is seen to decrease to ~ 0 for $\Delta N > 2$ for the peaks, whereas for valleys the correlation remains high, $\overline{C(0,3)} \sim 0.5$. This is also seen in Fig. 4(c) for $C(0, \Delta N)$ versus $\Delta N$ where peak-peak cross-correlations are seen to decrease faster than both valley-valley and peak-valley cross-correlations. The enhanced cross-correlation for valleys reflects the fact that, unlike resonant tunneling on peaks, elastic cotunneling relies on contributions from ~ $E/\Delta$ levels. In moving from one valley to the next, only one of the $E/\Delta$ levels is different, hence the similarity among neighbors.

We thank I. Aleiner and L. Glazman for many valuable discussions. Work at Stanford support in part by the Office of Naval Research YIP program, the Army Research Office, the NSF-NYI program, and the A.P. Sloan Foundation. Work at UCSB supported by the AFOSR and QUEST.

References

1. *Single Charge Tunneling , Proceedings of a NATO Advanced Study Institute,* edited by H. Grabert and M. H. Devoret (Plenum, New York, USA, 1992).

2. D. V. Averin and K. K. Likharev, in *Mesoscopic Phenomena in Solid* edited by B. L. Altshuler, P. A. Lee, and R. A. Webb (Elsevier, Amsterdam, 1991).

3. *Mesoscopic Phenomena in Solids*, edited by B. L. Altshuler, P. A. Lee, and R. Webb (North-Holland, Amsterdam, 1991).

4. C. M. Marcus, *et al.*, Phys. Rev. Lett. **69**, 506 (1992); M. W. Keller, O. Millo, A. Mittal, and D. E. Prober, Surf. Sci. **305**, 501 (1994); U. Sivan *et al.*, Europhys. Lett. **25**, 605 (1994); J. P. Bird,*et al.*, Phys. Rev. B **50**, 18678 (1994).

5. B. L. Altshuler and B. I. Shklovskii, Sov. Phys. JETP **64**, 127 (1986); B. L. Altshuler and B. D. Simons, in *Mesoscopic Quantum Physics*, edited by E. Akkermans, G. Montambaux, J.-L. Pichard, and J. Zinn-Justin (Elsevier, Amsterdam, 1995).

6. C. W. J. Beenakker, Phys. Rev. Lett. **70**, 1155 (1993); C. W. J. Beenakker, cond-mat /9612179 (1996).

7. R. A. Jalabert, H. U. Baranger, and A. D. Stone, Phys. Rev. Lett. **65**, 2442 (1990).

8. R. A. Jalabert, J.-L. Pichard, and C. W. J. Beenakker, Europhys. Lett. **27**, 255 (1994); H. U. Baranger and P. A. Mello, Phys. Rev. Lett. **73**, 142 (1994).

9. J. A. Folk *et al.*, Phys. Rev. Lett. **76**, 1699 (1996).

10. A. M. Chang, H. U. Baranger, L. N. Pfeiffer, K. W. West, and T. Y. Chang, Phys. Rev. Lett. **76**, 1695 (1996).

11. R. A. Jalabert, A. D. Stone, and Y. Alhassid, Phys. Rev. Lett. **68**, 3468 (1992).

12. Y. Alhassid and H. Attias, Phys. Rev. Lett. **76**, 1711 (1996).

13. H. Bruus, C. H. Lewenkopf, and E. R. Mucciolo, Phys. Rev. B **53**, 9968 (1996).

14. D. V. Averin and A. A. Odintsov, Phys. Lett. A **140**, 251 (1989).

15. L. I. Glazman and K. A. Matveev, Sov. Phys. JETP Lett. **51**, 484 (1990).

16. D. V. Averin and Y. V. Nazarov, Phys. Rev. Lett. **65**, 2446 (1990).

17. D. C. Glattli, C. Pasquier, U. Meirav, F. I. B. Williams, Y. Jin, and B. Etienne, Z. Phys. B **85**, 375 (1991).

18. I. L. Aleiner and L. I. Glazman, Phys. Rev. Lett. **77**, 2057 (1996).

19. E. Doron, U. Smilansky, and A. Frenkel, Physica D **50**, 367 (1991).

20. C. M. Marcus, R. M. Westervelt, P. F. Hopkins, and A. C. Gossard, Phys. Rev. B **48**, 2460 (1993).

21. C. W. J. Beenakker, Phys. Rev. B **44**, 1646 (1991).

22. V. N. Prigodin, K. B. Efetov, and S. Iida, Phys. Rev. Lett. **71**, 1230 (1993).

23. H. U. Baranger, R. A. Jalabert, and A. D. Stone, Phys. Rev. Lett. **70**, 3876 (1993).

24. A. M. Chang, H. U. Baranger, L. N. Pfeiffer, and K. W. West, Phys. Rev. Lett. **73**, 2111 (1994); M. J. Berry, J. A. Katine, C. M. Marcus, R. M. Westervelt, and A. C. Gossard, Surf. Sci. **305**, 495 (1994); I. H. Chan, R. M. Clarke, C. M. Marcus, K. Campman, and A. C. Gossard, Phys. Rev. Lett. **74**, 3876 (1995).

25. P. A. Lee and T. V. Ramakrishnan, Rev. Mod. Phys. **57,** 287 (1995).

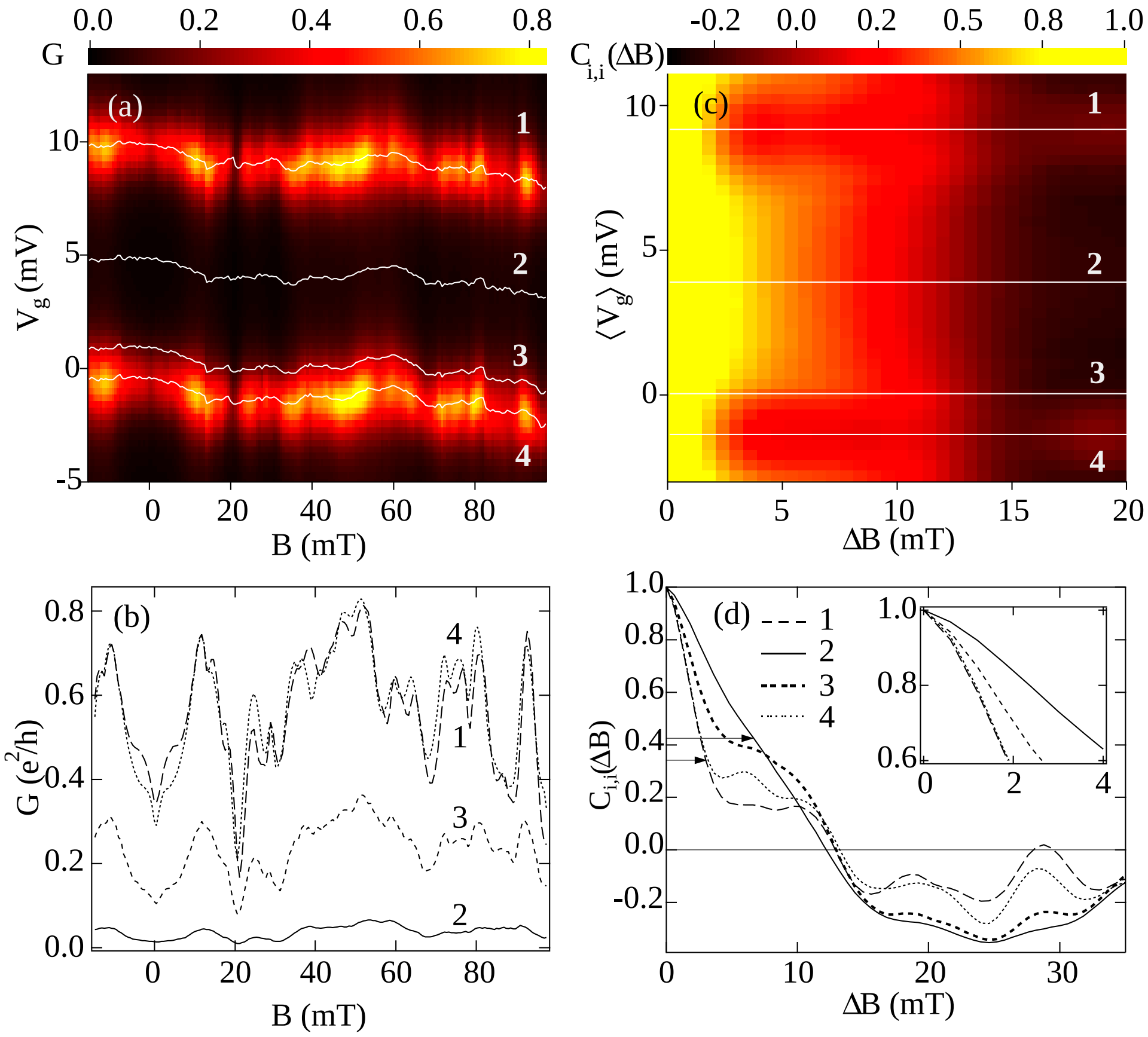

Fig. 1: (a,b) Conductance, $G$, as a function of magnetic field, $B$, and (a) as a function of gate voltage, $V_g$, and (b) for four particular paths. Paths used in (b) are shown as labeled white traces in (a). (c,d) Autocorrelations $C_{i,i}(\Delta B)$ of conductance fluctuations for (c) all interpolated paths as a function of average gate voltage along path, $\langle V_g \rangle$, and (d) for four traces. (d) Inset: $C_{i,i}(\Delta B)$ near $\Delta B = 0$.

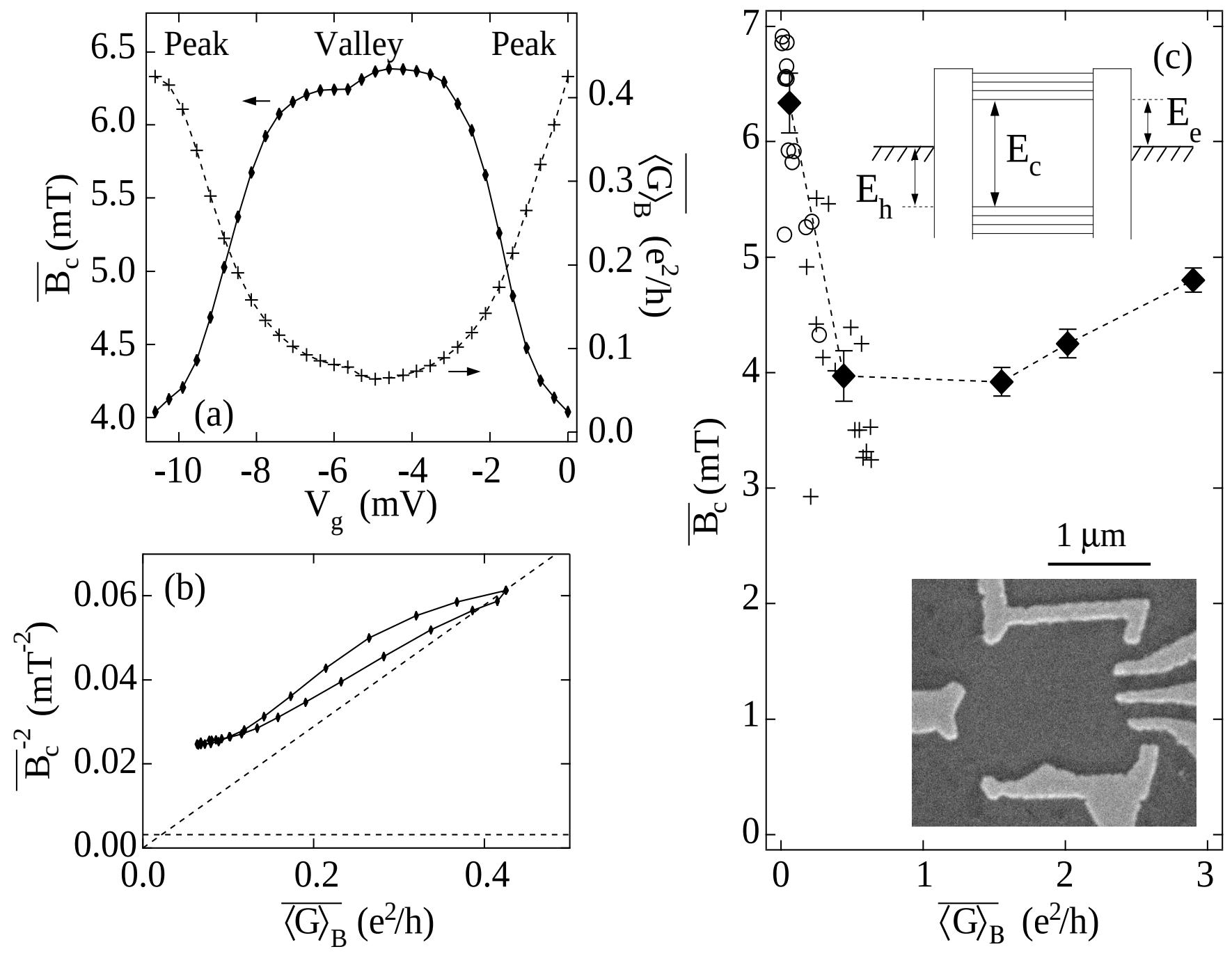

Fig. 2: (a) Ensemble-averaged characteristic field $\overline{B_c}$ (solid) and average conductance $\overline{\langle G\rangle_B}$ (dashed) across peak–valley–peak for ~14 independent data sets (dot 1), showing modulation of $\overline{B_c}$. (b) $\overline{B_c}^{-2}$ vs. $\overline{\langle G\rangle_B}$ for the same data. Diagonal line indicates $B_c{}^2\langle G\rangle$ = constant; horizontal line is saturation $\overline{B_c}^{-2} = (k\phi_0/A)^{-2}$ = 0.003 mT$^{-2}$ for $E_T < E$. (c) Average $\overline{B_c}$ for three open dot configurations and peak and valley conductances (diamonds), from data in (a). Unaveraged $B_c$ values for peaks (crosses) and valleys (circles) show spread in data. Open-dot $\overline{B_c}$ is modified (by ~ 5-15 %) to reflect changes in dot area upon opening leads. Top inset: Schematic energy diagram of a blockaded dot. Bottom inset: micrograph of dot 1.

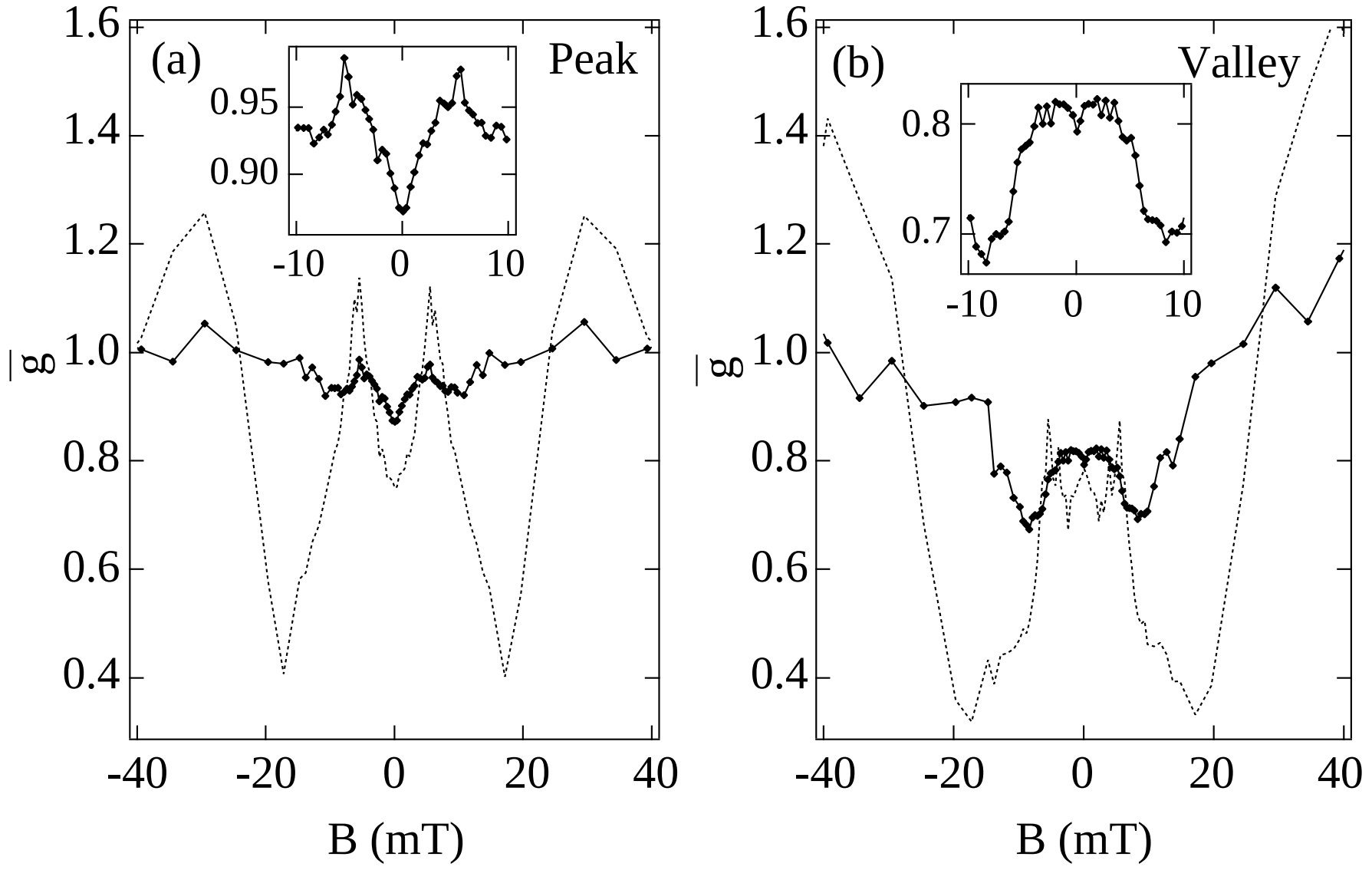

Fig. 3: Ensemble-averaged normalized conductance, $\bar{g}$, (solid) for 81 statistically independent (a) peaks and (b) valleys as a function magnetic field for dot 3. Peak conductance $\bar{g}_{peak}$ has a dip around $B = 0$ with a width $\sim B_c$ associated with the breaking of time-reversal symmetry. No dip around $B = 0$ is seen in $\bar{g}_{valley}$. Insets show the regions around B = 0. Unaveraged conductance (dashed) shows large fluctuations around average values.

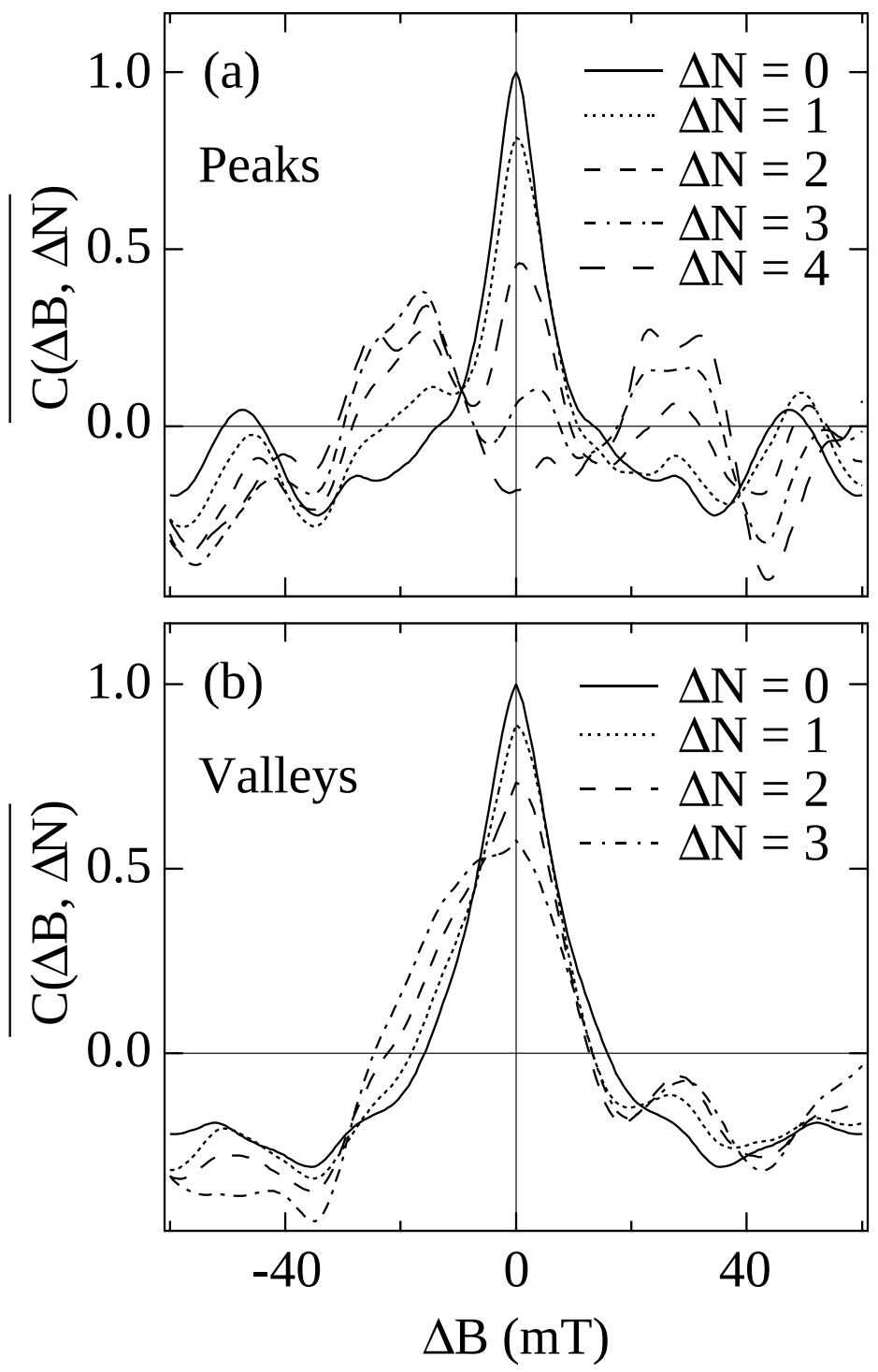

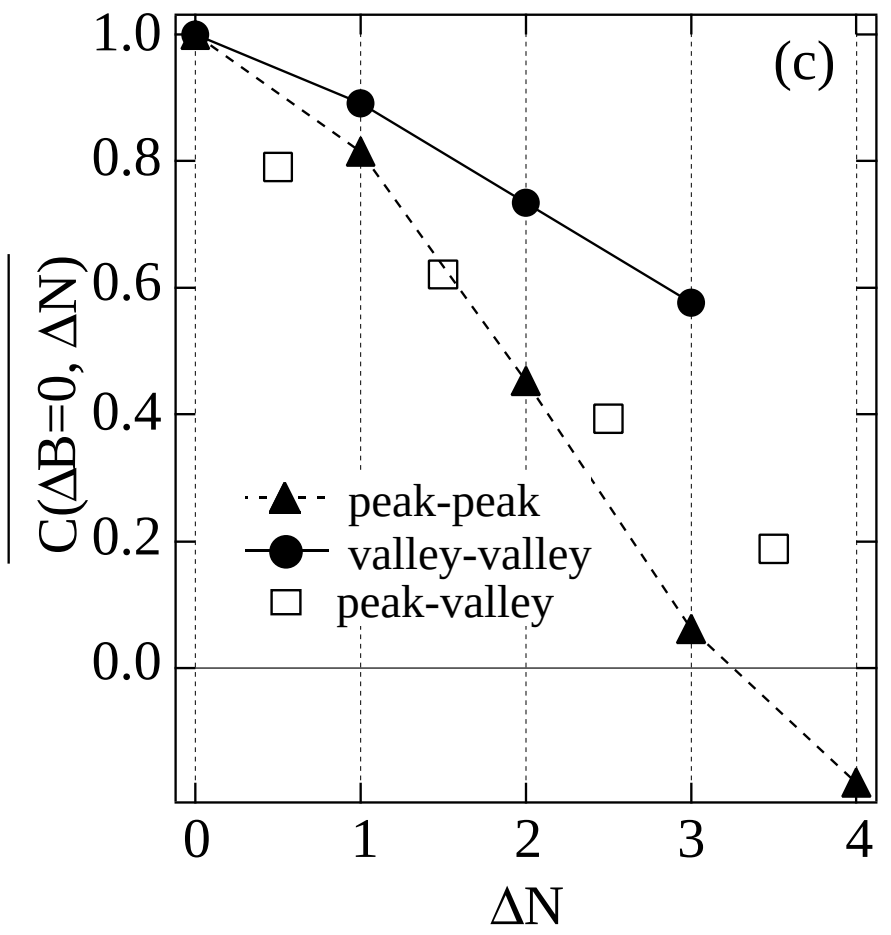

| parameter | dot 1 | dot 2 | dot 3 |
|---|---|---|---|
| $l$ (μm) | 8.4 | 8.4 | 4.3 |
| A (μm$^2$) | 0.81 | 0.45 | 0.57 |
| Δ (μeV) | 8.8 | 15.8 | 12.4 |
| $E_c$ (μeV) | 330 | 410 | 250 |
| $E_T$ (μeV) | 140 | 180 | 200 |

Fig. 4: Average cross-correlation of conductance fluctuations, $\overline{C(\Delta B, \Delta N)}$, for (a) peaks and (b) valleys. (c) Maximum cross-correlation at $\Delta B$ = 0, $\overline{C(0, \Delta N)}$, for peak-peak (triangles), valley-valley (circles) and peak-valley (squares), showing peak-peak correlations decrease more quickly than the others. (d) Device parameters for the three dots: mean free path ($l$), Area ($A$) based on ~ 150 nm depletion around gates, mean level spacing ($\Delta = 2\pi\hbar^2 / m^* A$), charging energy ($E_c = e^2/C$), and Thouless energy ($E_T = \hbar v_F / A^{1/2}$).